\documentstyle[sprocl,epsfig,wrapfig]{article}

\begin{document}

\pagestyle{empty}

\begin{flushleft}
\large
{SAGA-HE-158-00
\hfill January 13, 2000}  \\
\end{flushleft}
 
\vspace{1.6cm}
 
\begin{center}
 
{\LARGE 
       {\bf Parametrization of polarized parton distribution functions} \\

\vspace{1.5cm}
                  M. Hirai $^*$          \\
  \vspace{0.4cm}  Department of Physics \\
                  Saga University       \\
                  Saga 840-8502, Japan  \\
  \vspace{1.2cm}
}

{\Large
                  Talk at the RCNP-TMU Symposium on           \\
  \vspace{0.2cm}  ``Spins in Nuclear and Hadronic Reactions"  \\
  \vspace{0.5cm}  Tokyo, Japan, October 26 - 28, 1999         \\
  \vspace{0.05cm} (talk on Oct. 28, 1999)                     \\
}
\end{center}
 
\vspace{1.3cm}

\vfill
\noindent
{\rule{6.0cm}{0.1mm}} \\
 
\vspace{-0.3cm}
\noindent
$\ast$
Email: 98td25@edu.cc.saga-u.ac.jp. 
Information on his research is available at http://www-hs.phys.saga-u.ac.jp. \\

\vspace{+0.0cm}
\hfill
{\large to be published in proceedings}

\vfill\eject
\setcounter{page}{1}
\pagestyle{plain}

\title{Parametrization of polarized parton distribution functions}

\author{M. Hirai 
        \footnote{
        Information on his research is available at http://www-hs.phys.saga-u.ac.jp.}}
\address{Department of Physics, Saga University, Saga 840-8502, Japan \\
         E-mail: 98td25@edu.cc.saga-u.ac.jp}

\maketitle
\abstracts{
Polarized parton distribution functions are determined
by using asymmetry $A_1$ data from longitudinally polarized
deep inelastic scattering experiments. From our $\chi^2$ analysis,
polarized $u$-valence, $d$-valence, antiquark, and gluon distributions
are obtained. We propose one set of leading-order distributions and
two sets of next-to-leading-order ones as the longitudinally-polarized
parton distribution functions.
}

\section{Introduction}

After the EMC finding of a proton-spin issue, many
polarized deep-inelastic scattering (DIS) experiments
have been done on spin structure of the nucleon.
From these experimental data and theoretical studies,
we think that the nucleon spin is carried not only quarks
but also gluons and their angular momenta.
However, we do not have a clear idea even on the antiquark and gluon
contributions, which are difficult to be determined
by the present lepton-nucleon DIS data. The situation should become
clearer in the near future because RHIC-Spin experiments will
provide valuable information on these distributions.

We tried to determine the polarized parton distribution functions (PDFs)
by using existing spin asymmetry $A_1$ data for understanding the present situation 
and for suggesting the importance of future experimental
studies. The following discussions are based on the work in Ref. 1
with the members of the Asymmetry Analysis Collaboration (AAC).
In Sec. \ref{parton}, we explain how to calculate $A_1$ in terms of
the unpolarized and polarized parton distributions. Then, the actual
parametrization and $\chi^2$-analysis method are discussed in Sec. \ref{para}.
Our results are shown in Sec. \ref{results} and conclusions are given
in Sec. \ref{concl}.

\section{Parton model analysis of polarized DIS data}
\label{parton}

There are many measurements of the spin asymmetry $A_1$ for the proton, neutron, and deuteron.
To use these experimental data in our analysis, we express $A_1$ as
\begin{equation}
  A_1 (x, Q^2) \simeq \frac{g_1(x,Q^2)}{F_1(x,Q^2)} 
              =       \frac{2x [ 1+R(x,Q^2) ] }{F_2(x,Q^2)} g_1(x,Q^2),
\end{equation}
where $F_1$ and $F_2$ are unpolarized structure functions.
The function $R(x,Q^2)$ is given by $R(x,Q^2)=\sigma_L/\sigma_T$, 
where $\sigma_L$ and $\sigma_T$ are absorption cross sections of
longitudinal and transverse photons, and it is determined experimentally
in reasonably wide $Q^2$ and $x$ ranges in the SLAC experiment.~\cite{R1990}
The polarized structure function $g_1(x,Q^2)$ is expressed as
\small
\begin{equation}
  g_1(x,Q^2) = \frac{1}{2}\sum\limits_{i=1}^{n_f} e_{i}^2
     \bigg\{ \Delta C_q(x,\alpha_s) \otimes [ \Delta q_{i} (x,Q^2)
           + \Delta \bar{q}_{i} (x,Q^2) ] + \Delta C_g(x,\alpha_s)
    \otimes  \Delta g (x,Q^2) \bigg \},
\end{equation}
\normalsize
where $e_i$ is the electric charge of a quark, and the convolution $\otimes$ is
defined by 
$ f (x) \otimes g (x) = \int^{1}_{x} \frac{dy}{y} 
    f\left(\frac{x}{y} \right) g(y) .$
The distribution $\Delta q_i \equiv q^{\uparrow}_i-q^{\downarrow}_i$ represents
the difference between the number densities of quark with helicity parallel
to that of parent nucleon and with helicity anti-parallel.
The definitions of $\Delta \bar q_i$ and $\Delta g$ are the same.
$\Delta C_q$ and $\Delta C_g$ are the coefficient functions.
In discussing unpolarized reactions, 
the structure function $F_2$ is usually used rather than $F_1$,
and $F_2$ can be written in terms of unpolarized PDFs,
$q_i$, $\bar{q}_i$, and $g$, with coefficient functions
in the similar way to $g_1$.
In the next-to-leading-order (NLO) analysis, 
we choose the modified minimal subtraction ($\overline{\rm MS}$) scheme.

We provide the polarized parton distributions at $Q^2=1\,\rm{GeV}^2(\equiv Q_0^2)$.
Then, the distributions are evolved from $Q_0^2$ to experimental $Q^2$ points
by DGLAP equations. 
In our numerical analysis, we use a modified version of the program in Ref. 3,
where the evolution equations are solved by a brute-force method.

\section{Parametrization of polarized parton distributions}\label{para}
Now, we explain how the polarized parton distributions are parametrized.
The unpolarized PDFs $f_i(x, Q^2_0)$ and polarized PDFs $\Delta f_i(x, Q^2_0)$ 
are given at the initial scale $Q^2_0$.
Here, the subscript $i$ represents quark flavors and gluon.
In our analysis, we require the positivity condition of the PDFs 
in order to constrain the forms of the polarized PDFs.
Therefore, it is convenient to take the following functional form of the polarized PDFs at $Q^2_0$:
\begin{equation}
  \Delta f_i(x, Q^2_0) = A_i \, x^{\alpha_i} \,
                        (1 + \gamma_i \, x^{\lambda_i}) \, f_i(x, Q^2_0) .
  \label{eqn:PPDF2}
\end{equation}
The positivity condition is originated in a probabilistic interpretation of the parton densities:
the polarized PDFs should satisfy
$| \, \Delta f_i(x, Q^2_0) \, | \leq f_i(x, Q^2_0) .$
In our analysis, we simply require that this condition
should be satisfied not only in the leading-order (LO) and but also in the NLO at $Q_0^2$.
Thus, we have four parameters ($A_i$, $\alpha_i$, $\gamma_i$ and
$\lambda_i$) for each $i$.

In addition to the positivity condition, 
we assume the SU(3) flavor symmetry for the sea-quark distributions at $Q_0^2$
to reduce the number of free parameters.
Then, the first moments of $\Delta u_v (x)$ and $\Delta d_v (x)$, 
which are written as $\eta_{u_v}$ and $\eta_{d_v}$, 
can be described in terms of axial charges for octet baryon,
$F$ and $D$, measured in hyperon and neutron $\beta$-decays.
They are determined as
$F = 0.463 \pm 0.008$ and $D = 0.804 \pm 0.008$,
which lead to $\eta_{u_v}=0.926\pm 0.014$ and $\eta_{d_v}=-0.341 \pm 0.018$. 
In this way, we fix these first moments,
so that two parameters $A_{u_v}$ and $A_{d_v}$ are determined by these first moments 
and other parameter values. 
Thus, the remaining job is to determine the values of the following 14 parameters,
$A_{\bar q}, \ A_g, \ \alpha_i, \ \gamma_i, \ \lambda_i 
\ (i=u_v, \, d_v, \, \bar q, \, g)$,
by the $\chi^2$ analysis of the polarized DIS experimental data.

We determine the values of the 14 parameters by fitting the $A_1(x, Q^2)$ data for 
the proton from E130, E143, EMC, SMC, and HERMES, 
the neutron from E142, E154, and HERMES,
and the deuteron from E143, E155, and SMC.
We also use LO and NLO GRV parametrizations for the unpolarized PDFs \cite{GRV98} 
and the SLAC measurement of $R(x, Q^2)$.~\cite{R1990}
Then, the best parametrization is obtained by minimizing
\begin{equation}
\chi^2=\sum \frac{(A_1^{\rm data}(x,Q^2)-A_1^{\rm calc}(x,Q^2) )^2}
                 {(\Delta A_1^{\rm data}(x,Q^2) )^2},
\end{equation} 
where $\Delta A_1^{\rm data}$ represents the error on the experimental data including both
systematic and statistical errors.
In evolving the distribution functions with $Q^2$, we neglect charm-quark contributions
to $A_1(x,Q^2)$ and take the flavor number $N_f=3$,
because the contribution is very small in a few $Q^2$ region where most experimental data exist.
To be consistent with the unpolarized, we use the same scale parameters as the GRV,~\cite{GRV98}
$\Lambda_{\rm QCD}^{(3)}=\ 204 \ {\rm MeV}$ at LO and
$\Lambda_{\rm QCD}^{(3)}=\ 299 \ {\rm MeV}$ at NLO.

\section{Results}\label{results}
\begin{wrapfigure}{ri}{5.8cm}
  \vspace{-11mm}
  \begin{center}
    \epsfig{file=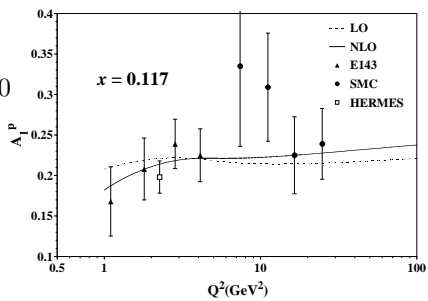,height=3.9cm}
  \end{center}
  \vspace{-3mm}
  \caption{Theoretical of spin asymmetries $A_1$ for the proton 
           are compared with the experimental data at $x \approx 0.117$. (from Ref. 1)}
  \label{fig:asym-1}
\end{wrapfigure}
We got minimum $\chi^2$:
$\chi^2$/d.o.f=322.6/360 for the LO and 
$\chi^2$/d.o.f=300.4/360 for the NLO.
The difference between the LO and NLO $\chi^2$ values is about 7\%,
which indicates the importance the NLO analysis.

We show the $Q^2$ dependence of spin asymmetry $A_1$ for the proton in Fig. \ref{fig:asym-1}.
This figure indicates that the NLO effects become larger in the small $Q^2$ region
and that there is strong $Q^2$ dependence especially in the small $Q^2$ region.
It is not right to assume $Q^2$ independence of the spin asymmetry $A_1$ 
in obtaining $g_1$,
so that, we have to be careful using PQCD in this region.

\begin{wrapfigure}{ri}{5.8cm}
  \vspace{-1mm}
  \begin{center}
    \epsfig{file=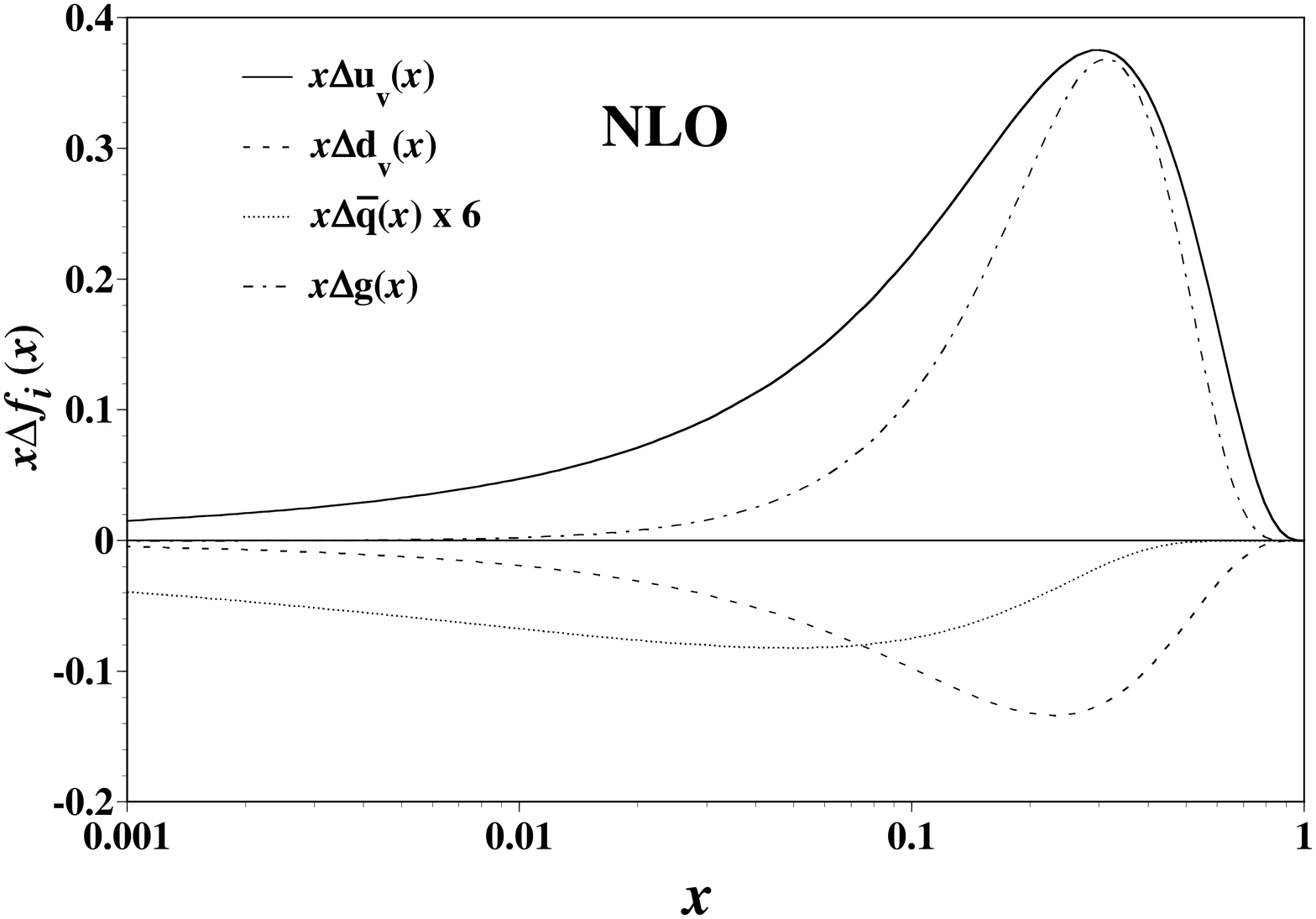,height=3.9cm}   
  \end{center}
  \vspace{-3mm}
    \caption{Obtained NLO polarized parton distributions at $Q^2=1$ GeV$^2$. (from Ref. 1)}
  \label{fig:df}
\end{wrapfigure}
\begin{wrapfigure}{ri}{5.5cm}
  \vspace{-2mm}
  \begin{center}
    \epsfig{file=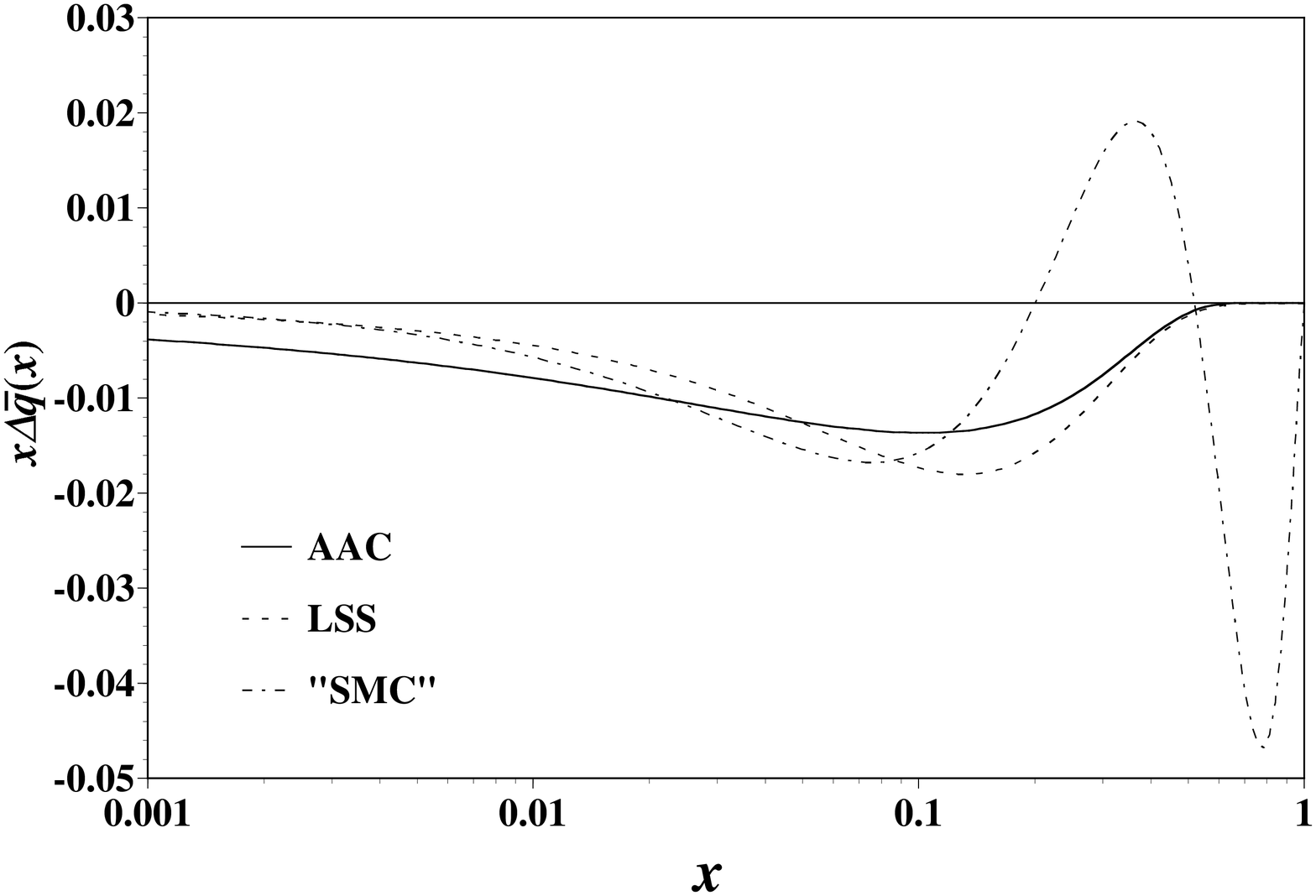,height=3.8cm}
  \end{center}
  \vspace{-4mm}
  \caption{Antiquark distributions of recent parametrizations ``SMC", LSS and  AAC 
           at $Q^2$=1 GeV$^2$. (from Ref. 1)}
  \label{fig:dqbar}
\end{wrapfigure}

Next, we show the polarized parton distributions for the NLO in Fig. \ref{fig:df}.
As the figure indicates, we obtain negative polarization for $\Delta \bar{q}(x)$, 
and large positive polarization for $\Delta g(x)$.
The first moment for $\Delta u_v (x)$ is fixed at the positive value 
and the one for $\Delta d_v (x)$ is fixed at the negative value, 
so that the obtained distributions $\Delta u_v(x)$ and $\Delta d_v(x)$
become positive and negative, respectively.
Similar results are obtained in the LO distributions.~\cite{AAC}
There are slight differences between the LO and NLO distributions $\Delta d_v (x)$.
However, the differences are large between the LO and NLO gluon distributions 
in the wide $x$ region.
It is caused by the gluon contribution through the coefficient function.
%
We calculate the quark spin content by using the obtained LO and NLO
distributions. It is given by 
$\Delta \Sigma =\eta_{u_v}+\eta_{d_v}+6\,\eta_{\bar{q}}$, where
$\eta_{\bar{q}}$ is the first moment of $\Delta \bar{q}(x)$.
Because $\eta_{u_v}$ and $\eta_{d_v}$ are fixed, only
$\eta_{\bar{q}}$ affects the spin content in the different analyses.
The LO and NLO moments are $\eta_{\bar{q}}= -0.064$ and $-0.089$,
so that the spin content becomes $\Delta \Sigma= 0.201$ and $0.051$,
respectively.~\cite{AAC} The NLO spin content ($\Delta \Sigma=0.051$)
is significantly smaller than other analysis results. For example,
the recent SMC and Leader-Sidrov-Stamenov (LSS) parametrizations
\cite{SMCfit,LSS} obtained $\Delta\Sigma=$0.19 and 0.28 at $Q^2$=1 GeV$^2$.
In order to investigate the reason for the small $\Delta \Sigma$ in our analysis, 
we show each antiquark distribution in Fig. \ref{fig:dqbar}.

The NLO antiquark distributions of the SMC, LSS, and AAC analyses
are calculated at $Q^2=$1 GeV$^2$. 
Because the antiquark distribution is not directly given in the SMC
analysis, we may call it as a transformed SMC (``SMC") distribution.
It is calculated by transforming the published distributions by the SMC.
All the distributions agree in principle in this region ($0.01<x<0.1$) 
where accurate experimental data exist and the antiquark distribution plays an important role.
On the other hand, it is clear that our distribution does not fall off rapidly 
as $x \rightarrow 0$ in comparison with the others.
This is the reason why our NLO spin content is significantly smaller.
In fact, we obtained the parameter $\alpha$ for the antiquark distribution
as $\alpha_{\bar q} (NLO) = 0.32 \pm 0.22$,
which controlled the small-$x$ behavior of $\Delta \bar{q}(x)$.
However, the large error of the parameter $\alpha_{\bar q}$
suggests that the small-$x$ part of $\Delta \bar{q}(x)$
cannot be fixed by the existing data. 
Actually, there is no data in the small-$x$ region $(x < 0.04)$.
Therefore, we had better consider to constrain the parameter $\alpha_{\bar q}$
by theoretical ideas.
We discuss such possibilities by using the Regge theory and the perturbative QCD. 

According to the Regge model, the small-$x$ behavior of $g_1$
is suggested as $g_1 (x)  \sim  x^{-\alpha}$ with $\alpha=-0.5 \sim 0.0$.~\cite{regge} 
Therefore, we expect $\Delta \bar{q}(x) \sim  x^{0.0 \sim 0.5}$ as $x\rightarrow 0$.
Because the parametrized function is given by $\Delta \bar{q}(x)/\bar{q}(x)$,
we should find out the small-$x$ behavior of the unpolarized distribution.
The GRV distribution has the property $x \, \bar q \sim x^{-0.14}$ at $Q^2$=1 GeV$^2$
according to our numerical analysis,
the Regge prediction becomes $\alpha_{\bar q}^{Regge} = 1.1 \sim 1.6$,
if the theory is applied at $Q^2=1$ GeV$^2$.
The perturbative QCD could also suggest the small-$x$ behavior.
If we can assume that the singlet-quark and gluon distributions are constants
as $x\rightarrow 0$ at certain $Q^2$ ($\equiv Q_1^2$),
their singular behavior is predicted from the evolution equations.~\cite{lr-sum}
The singlet distribution behaves like 
$\Delta \Sigma (x) \sim x^\alpha$ as $\alpha ={-0.12 \sim -0.09}$,
if we choose the evolution range from $Q_1^2=0.3 \, \sim \, 0.5$ GeV$^2$ to $Q^2=1$ GeV$^2$.
Therefore, the perturbative QCD with the assumption of the above $Q_1^2$ range suggests
$\alpha_{\bar q}^{pQCD} = 1.0$.

\begin{wrapfigure}{ri}{5.5cm}
  \vspace{-1.5mm}
  \begin{center}
     \epsfig{file=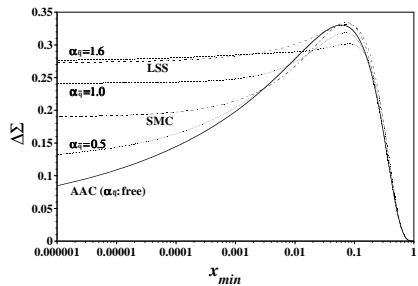,height=3.7cm}
  \end{center}
  \vspace{-4mm}
  \caption{$x_{min}$ dependence of $\Delta \Sigma$ at $Q^2=1\,\rm{GeV}^2$. (from Ref. 1)}
  \label{fig:dsigma}
\end{wrapfigure}
The Regge theory and perturbative QCD suggest the range 
$\alpha_{\bar q}=1.0 \sim 1.6$, so that we try the NLO $\chi^2$ analyses by
fixing the parameter at $\alpha_{\bar q}=$0.5, 1.0, and 1.6.
The last two values are in the theoretical prediction range,
and the first one is simply taken as a slightly singular distribution.
The obtained minimum $\chi^2$ values are larger than the NLO fit ($\chi^2$=300.4)
by 0.1, 1.8, and 7.7\%, and the first moments are
$\Delta \Sigma$=0.123, 0.241 and 0.276
for $\alpha_{\bar q}=$0.5, 1.0, and 1.6, respectively.
The small-$x$ falloff for larger $\alpha_{\bar q}$ changes
the $\eta_{\bar{q}}$ and $\Delta \Sigma$ significantly. 
We show the spin content in the region between $x_{min}$ and 1
by calculating  $\Delta \Sigma=\int _{x_{min}}^1 \Delta \Sigma (x)dx$
in Fig. \ref{fig:dsigma}.
Because the LSS and SMC distributions are less singular functions of $x$,
their spin contents saturate even at $x_{min}=10^{-4}$ although the $\Delta \Sigma$
of our NLO result with free $\alpha_{\bar q}$ still decreases in this region.
If the parameter $\alpha_{\bar q}$ is taken in the perturbative QCD
and Regge theory prediction range,
the calculated spin content is within the usually quoted values $\Delta\Sigma=0.1\sim 0.3$.
In this sense, our results are not inconsistent with the previous analyses. 
Our results indicate that the spin content cannot be determined uniquely,
because the accurate experimental data are not available in small $x$ region.
The obtained $\chi^2$ value suggests that the $\alpha_{\bar q}$=1.0
solution could be also taken as one of the good fits. The $\alpha_{\bar q}$=0.5
distributions are almost the same as the ones in the free-$\alpha_{\bar{q}}$ NLO analysis,
so that it is redundant to propose it as one of our good fits.

From our analyses, we propose the LO distributions, the NLO ones 
with free $\alpha_{\bar q}$ (NLO-1), 
and those with fixed $\alpha_{\bar q}$=1.0 (NLO-2) as the longitudinally-polarized
parton distributions of the AAC analyses. Useful functional
forms are given at $Q^2$=1 GeV$^2$ in Appendix B of Ref. 1
for practical applications. 

\vspace{-3mm}
\section{Conclusions}\label{concl}
From the LO and NLO $\chi^2$ analyses,
we obtained good fits to the experimental data. 
Because the NLO $\chi^2$ is significantly smaller than that of LO, 
the NLO analysis should be necessarily used in the parametrization studies.
It is particularly important for extracting information on $\Delta g$. 
However, the polarized antiquark and gluon distribution cannot be uniquely determined
by the present DIS data. 
We provide the optimum LO and NLO distributions at $Q^2=1\,\rm{GeV}^2$ from our numerical analyses.

\vspace{-2mm}
\section*{Acknowledgments}
M.H. was supported as a Research Fellow of the Japan Society for the Promotion of Science,
and he would like to thank S. Kumano and M. Miyama for reading this manuscript.
This talk is based on the work with Y. Goto, N. Hayashi, M. Hirai,
H. Horikawa, S. Kumano, M. Miyama, T. Morii, N. Saito, T.-A. Shibata,
E. Taniguchi, and T. Yamanishi.~\cite{AAC}


\end{document}